\documentclass[aps,pra,singlecolumn,superscriptaddress,notitlepage]{revtex4-1}

\usepackage{graphicx}
\usepackage{amsmath}
\usepackage{amssymb}
\usepackage{epstopdf}
\usepackage{braket}
\usepackage{array}
\usepackage[dvipsnames]{xcolor}
\usepackage{color}
\usepackage{cancel}
\usepackage{multirow}
\usepackage{comment}
\usepackage[normalem]{ulem}
\usepackage[colorlinks,linkcolor=blue,urlcolor=blue,citecolor=blue,anchorcolor=blue]{hyperref}
\usepackage{enumitem}

\begin{document}

\title{Single-photon quantum hardware: towards scalable photonic quantum technology with a quantum advantage}

\author{Ravitej Uppu}
\affiliation{Center for Hybrid Quantum Networks (Hy-Q), Niels Bohr Institute, University of Copenhagen, Blegdamsvej 17, DK-2100 Copenhagen, Denmark}
\author{Leonardo Midolo}
\affiliation{Center for Hybrid Quantum Networks (Hy-Q), Niels Bohr Institute, University of Copenhagen, Blegdamsvej 17, DK-2100 Copenhagen, Denmark}
\author{Xiaoyan Zhou}
\affiliation{Center for Hybrid Quantum Networks (Hy-Q), Niels Bohr Institute, University of Copenhagen, Blegdamsvej 17, DK-2100 Copenhagen, Denmark}
\author{Jacques Carolan}
\affiliation{Center for Hybrid Quantum Networks (Hy-Q), Niels Bohr Institute, University of Copenhagen, Blegdamsvej 17, DK-2100 Copenhagen, Denmark}
\author{Peter Lodahl}
\email{lodahl@nbi.ku.dk}
\affiliation{Center for Hybrid Quantum Networks (Hy-Q), Niels Bohr Institute, University of Copenhagen, Blegdamsvej 17, DK-2100 Copenhagen, Denmark}

\begin{abstract}
The scaling up of quantum hardware is the fundamental challenge ahead in order to realize the disruptive potential of quantum technology in information science.
Among the plethora of hardware platforms, photonics stands out by offering a modular approach, where the main challenge is to construct sufficiently high-quality building blocks and develop methods to efficiently interface them.
Importantly, the subsequent scaling-up will make full use of the mature integrated photonic technology provided by photonic foundry infrastructure to produce small foot-print quantum processors of immense complexity.
A fully coherent and deterministic photon-emitter interface is a key enabler of quantum photonics, and can today be realized with solid-state quantum emitters with specifications reaching the quantitative benchmark referred to as Quantum Advantage. 
This light-matter interaction primer realizes a range of quantum photonic resources and functionalities, including on-demand single-photon and multi-photon entanglement sources, and photon-photon nonlinear quantum gates.
We will present the current state-of-the-art in single-photon quantum hardware and the main photonic building blocks required in order to scale up.
Furthermore, we will point out specific promising applications of the hardware building blocks within quantum communication and photonic quantum computing, laying out the road ahead for quantum photonics applications that could offer genuine quantum advantage.
\end{abstract}

\maketitle


\section*{Photonic quantum information processing}
\begin{figure}
    \centering
    \includegraphics[width=150mm]{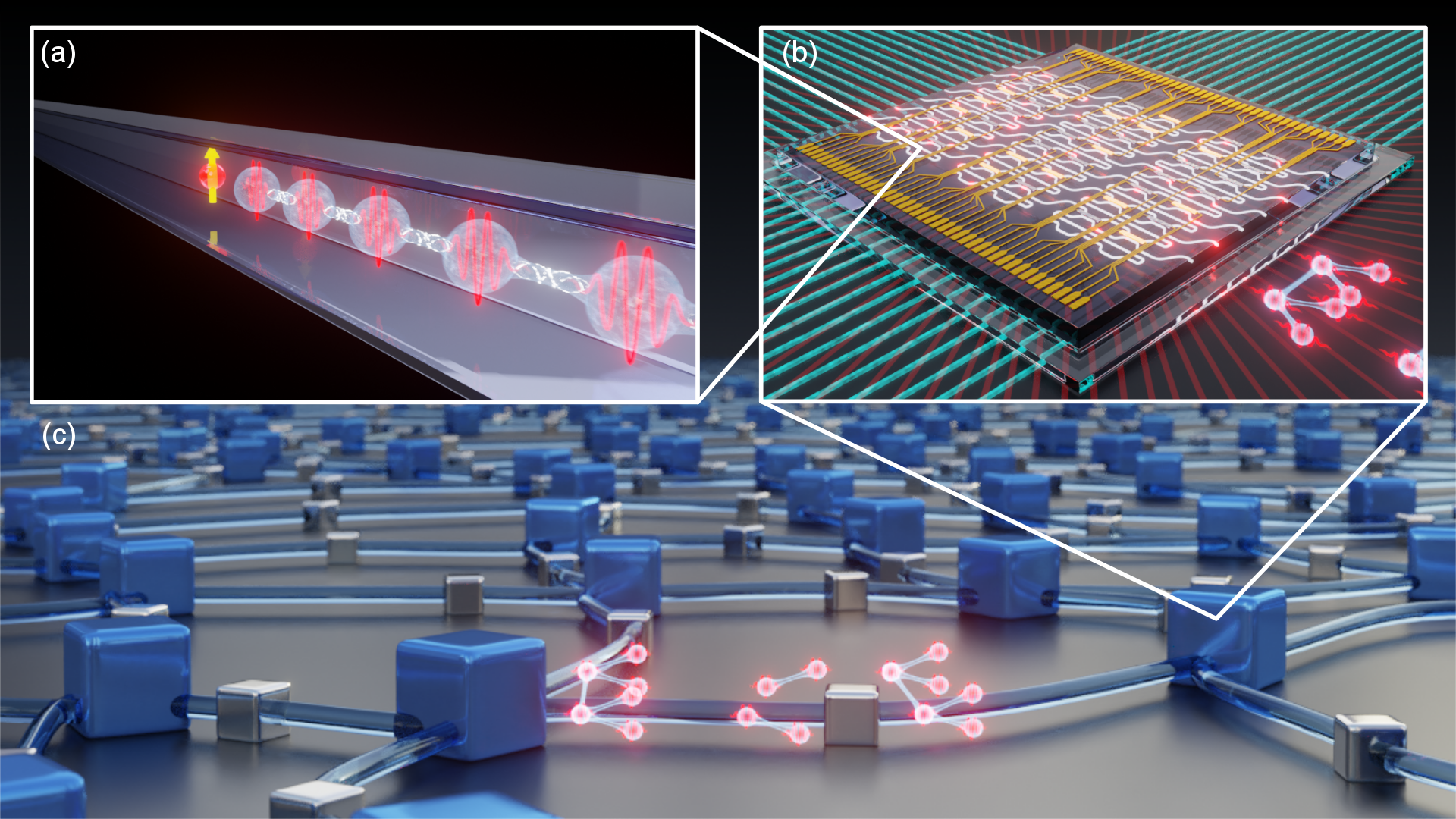}
    \caption{Scalable and modular quantum photonic technologies based on deterministic single-photon quantum hardware. (a) Illustration of a deterministic photon-emitter interface in a nanophotonic waveguide. (b) Illustration of PIC configured for synthesizing multi-photon entangled states. (c) Artist view of advanced photonic quantum networks where communication between network links is facilitated by multi-photon entangled pulses.}
    \label{fig:fig1}
\end{figure}

Photonic quantum hardware is rapidly developing where the benefits of advanced photonic chip technology are exploited. 
Quantum photonics therefore appears as a front runner in quantum technology where real-world applications emerge early. 
Photonics is indispensable in quantum communication where pulses of light are carriers of quantum information through optical fibers.
Figure \ref{fig:fig1} illustrates the concepts covered in the present manuscript.
Our point of departure is the availability of efficient photon-emitter interfaces providing deterministic and fully quantum coherent light-matter interaction, cf. Fig. \ref{fig:fig1}(a), providing a basis for deterministic single- and multi-photon sources.
Subsequently, photonic integrated circuit (PIC) technology can be exploited for scaling up, e.g., by processing many photonic qubits or for synthesizing advanced photonic resources, cf. Fig. \ref{fig:fig1}(b). In contrast, other probabilistic sources require heralding for scaling up which results in excess resource overhead \cite{Rudolph2017APLphoton}. 
These PIC-based quantum processors could potentially be implemented in applications, cf. \ref{fig:fig1}(c), and we will outline specific architectures tailored to quantum-dot (QD) single-photon hardware within quantum communication and photonic quantum computing.

\section{Deterministic and quantum coherent photon-emitter interfaces}
\begin{figure}
    \centering
    \includegraphics[width=160mm]{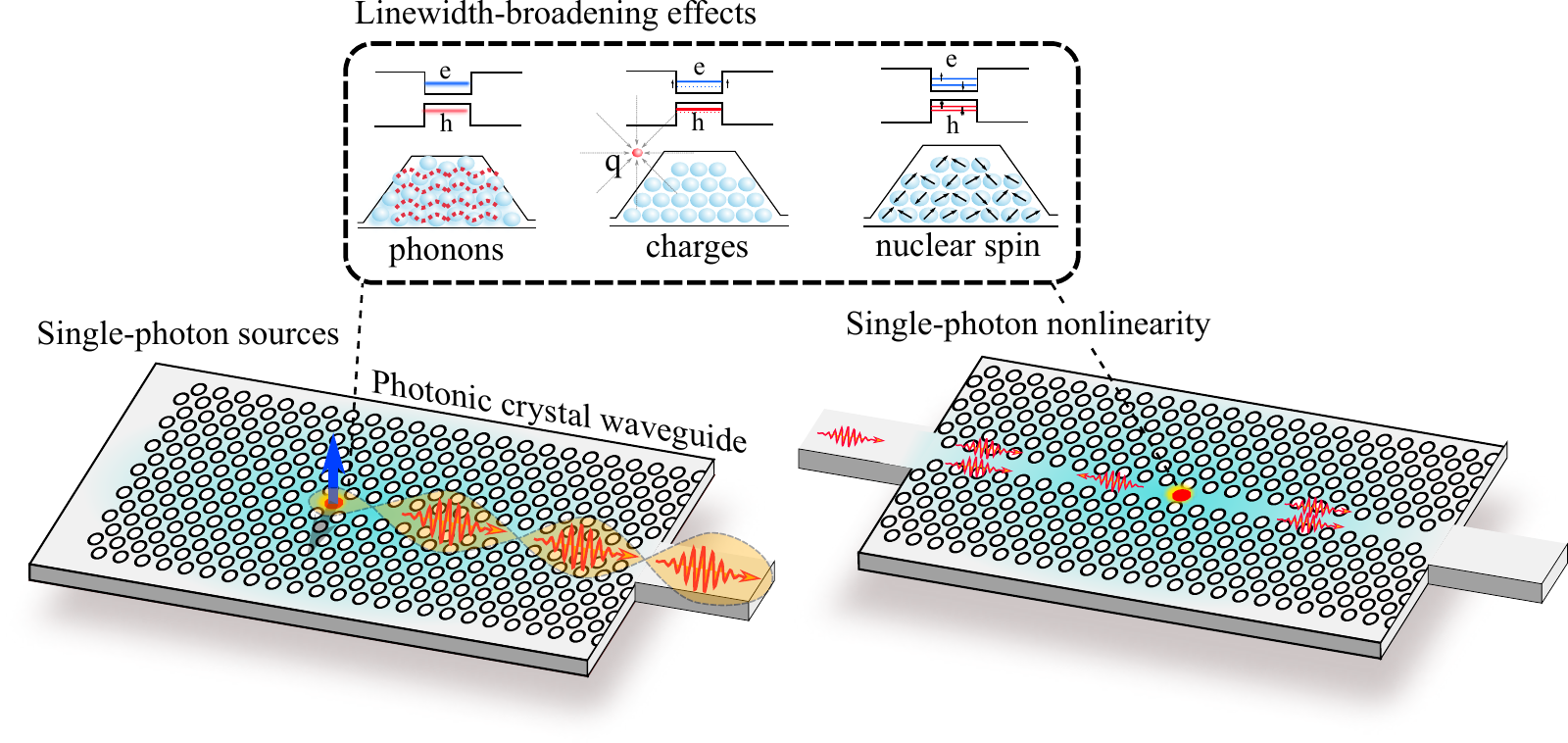}
    \caption{Illustration of a deterministic photon-emitter interface for the exemplary case of a QD in a planar nanophotonic waveguide. The concepts are general and apply to other types of emitters and cavity/waveguide implementations as well. The relevant deteriorating  decoherence processes for QDs leading to linewidth broadening are shown, including coupling to phonons, charges, and a fluctuating nuclear spin bath that additionally introduces decoherence of the electronic spin degrees of freedom. The devices can be operated either as a single-photon or entanglement source (left illustration) or as a giant nonlinearity that operates at the single-photon level (right illustration).}
    \label{fig:emitterphoton}
\end{figure}

A single quantum emitter, e.g., an atom, ion or solid-state emitter, constitutes the fundamental quantum interface between light and matter. 
It couples a single excitation of light (the photon) to a single atomic excitation. 
The coupling is usually weak and any incoherent dephasing processes may deteriorate the inherent quantum character of the interaction. 
Both challenges have recently been overcome by using quantum emitters in photonic nanostructures after implementing careful shielding of external noise, cf. Fig. \ref{fig:emitterphoton}.
Different quantum emitters are considered at optical frequencies, including QDs, atoms, vacancy centers in diamond, molecules, or excitons in two-dimensional materials \cite{Aharonovich2016np}. 
Furthermore, the underlying physics applies as well to superconducting qubits in resonators and waveguides \cite{Blais2020arXiv} although their operation at microwave frequencies precludes the applications in quantum communication. 
In order to present precise figures-of-merit and benchmarks that are essential for projecting out to the proposed applications, we will focus on QDs in nanophotonic cavities and waveguides.
This platform has recently matured dramatically leading to the realization of a near-deterministic and coherent photon-emitter interface \cite{senellart2017natnano,Wang2019natphton,uppu2020scienceadv,tomm2021nn}.

A QD in a single-mode waveguide or nanocavity is a prototypical implementation of a deterministic photon-emitter interface.
Figure \ref{fig:emitterphoton} illustrates the case of, e.g., an InAs QD placed in a photonic-crystal waveguide where the QD is positioned such that it couples efficiently to the spatially varying fundamental waveguide mode. 
The QD exhibits two linear orthogonal transition dipoles, and the waveguide can be aligned with one dipole  maximally coupled to the fundamental waveguide mode, while the other dipole is suppressed.
The leakage to other modes can be suppressed in cavities and waveguides, which is quantified by the $\beta$-factor specifying the probability that the QD emits into the desired mode. 
Near-unity $\beta$-factors are routinely achieved in nanophotonics \cite{lodahl2015rmp}, however for quantum applications all quantum decoherence processes must be suppressed as well. 
To this end, a relevant figure-of-merit is the degree of indistinguishability (ID) of subsequently emitted photons, and for QDs ID above 95 \% was reported over extended time scales to produce more than hundred indistinguishable photons \cite{uppu2020scienceadv,tomm2021nn}. 
The photon-emitter interface can be operated as an on-demand source of single photons by resonantly exciting the QD that subsequently emits photons into the waveguide. 
In another configuration, resonant photons are launched into the waveguide and the QD serves as a giant non-linearity that introduces strong correlations between individual photons. 
These two cases are illustrated in Fig. \ref{fig:emitterphoton} and highlight the versatility of the approach.

Realizing highly coherent emitters has been an outstanding challenge for solid-state emitters and requires identifying and combating all noise processes affecting photon emission. 
For QDs, the relevant decoherence processes are sketched in Fig. \ref{fig:emitterphoton}. 
They include phononic broadening due to a finite temperature, charge noise from electric charges in the vicinity of the QD, and spin noise from the coupling of the electron spin to the randomly-oriented nuclear spins of the atoms making up the QD. 
Remarkably, charge noise can be fully suppressed in epitaxially grown and electrically-contacted samples  and nuclear spin noise only leads to minor broadening effects  \cite{Kuhlmann2013NatPhys}. 
Consequently, transform-limited QD emission has been demonstrated \cite{kuhlmann2015nc}, which subsequently was realized also in high $\beta$-factor nanophotonic waveguides \cite{pedersen2020ACSphoton}. 
In this manner, a coherent and deterministic photon-emitter interface is realized. 

To realize high-fidelity quantum operation, as required for advanced quantum applications, even minor decoherence contributions need  to be accounted for.
Phonon scattering remains the fundamental decoherence mechanism that contributes even at cryogenic temperatures, which limits the reported values of ID to high nineties. 
Going beyond, an experimentally feasible strategy has been identified for increasing ID above $99 \%$ through phonon damping by proper clamping of the nanostructures \cite{dreessen2018qst}. 
An alternative strategy applies strong Purcell enhancement to increase the emission rate relative to the decoherence rate  \cite{Santori2002Nature}. 
It has already been established that state-of-the-art QD single-photon sources suffice for realizing Quantum Advantage in a boson sampling quantum-simulation algorithm requiring about 50 high-quality photons \cite{uppu2020scienceadv}, see Sec. II for further discussion. 
Further experiments will likely show that these sources can emit thousands if not millions of highly indistinguishable photons since the photon emission is much faster (typically 100 ps) than slow residual drift processes (typically milliseconds). 
It will be exciting to see in the future whether this massive photonic quantum resource delivered by just a single QD can be a key enabler in advanced quantum-information processing applications.
This is intimately connected to how efficient the generated string of photons can be coupled, switched, and processed, which are topics covered in the following section of the present manuscript.  

Electrically contacted QDs have the additional benefit that various charge states can be deterministically prepared leading to diverse opportunities. 
Loading the QD with a single electron or hole introduces a two-fold metastable ground state corresponding to spin up/down relative to an external magnetic field serving as a quantum memory of the system. 
The spin coherence is limited, however, due to the coupling of the charge to the nuclear spin bath. 
The typical  spin dephasing time ($T_2^* $) is nanoseconds for single electrons that can be extended to hundred nanoseconds for hole spins \cite{huthmacher2018prb}. 
However, the spin coherence time ($T_2$), which is relevant in protocols where spin-echo refocusing is implemented, can reach the level of microseconds \cite{Stockill2016natcomm}. 
Importantly, since emission is rapid, a QD can emit many photons within the spin coherence time, which is essential for the scalability of advanced multi-photon entanglement sources, as will be described in a later section.

Multi-particle excitations provide further opportunities. 
Biexciton states consist of two electrons/holes in the conduction/valence bands of the QD and recombine through a cascaded two-photon process. 
The availability of two indistinguishable decay paths lead to the generation of polarization-entangled photons on-demand  \cite{Benson2000prl,liu2019natnano}. 
Coupling multiple QDs is another essential requirement, e.g., for the generation of advanced multi-dimensional entangled states. 
This can be achieved by exploiting coherent electronic tunnel coupling \cite{greilich2011np} or by optical dipole-dipole interaction possibly engineered by the photonic nanostructure \cite{Grim2019natmat}. 
QD inhomogeneities introduced during growth remains a major challenge in order to scale up from present-day few QD experiments to many. 
Importantly, most of the applications that are considered in the present manuscript require only a few and sometimes even just one QD. 
Nonetheless overcoming QD inhomogeneous broadening would constitute a major breakthrough and new selective-area growth methods may provide a path way to realize the required control on the atomic scale \cite{krizek2018prm}.


\begin{figure}
    \centering
    \includegraphics[width=0.5\columnwidth]{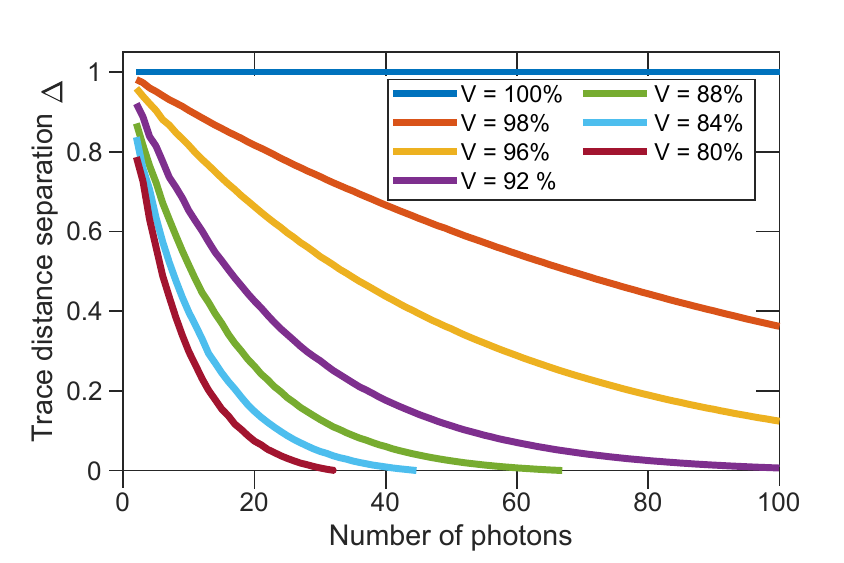}
    \caption{ Photon quality requirements for scaling up the boson sampling algorithm. The plot shows the trace distance separation $(\Delta)$ \cite{Shchesnovich2015pra} between a real boson sampler with a finite degree of ID (V) of the photons and a perfect boson sampler with $V = 100 \%$ as a function of the number of photons. $\Delta=0$ corresponds to the computationally easy case of distinguishable photon, while $\Delta>0$ is the regime where computational hardness appears. $V \geq 96\%$ has been realized with QDs over long strings of many photons \cite{uppu2020scienceadv,tomm2021nn} enabling boson sampling in the QA regime of $\approx 50$ photons with $\Delta>0.4$.}
    \label{fig:bosonsampling}
\end{figure}

\section{Quantum hardware enabling a quantum advantage}
What are the performance requirements of quantum hardware for carrying out tasks that are impossible with classical hardware? 
Obviously this question has no simple answer, since it would depend on the specific application targeted among the multitude of potential applications of quantum technology. 
Nonetheless this question is essential for the researchers developing the quantum hardware, since clear performance benchmarks are required for guiding future work. 
Despite the diversity of applications often the same physical parameters are relevant, since transformative quantum applications rely on similar physical principles, e.g., quantum-superposition states or multi-particle entanglement. 
In the main text, the essential physical parameters for photonic qubits are discussed. 

Quantitative benchmarking  requires zooming in on the precise application. 
In the context of quantum simulations/computing, a quantitative benchmark is defined that is generally referred to as Quantum Advantage (QA) \footnote{In the literature also the term Quantum Supremacy prevails. However, as has been discussed in a commentary (Palacios-Berraquero et al., \textit{Nature} \textbf{576}, 213 (2019)), the term Quantum Advantage captures the importance of this rather technical quantum hardware achievement while avoiding associations to the historical meaning of the term supremacy.}: QA signifies the threshold at which the accessible hardware can implement a specific quantum algorithm that cannot be realized on even the World's largest supercomputer \cite{preskill2012arx} since the classical algorithm scales exponentially with the size of the system being simulated. 
QA has been realized with superconducting qubits in 2019 by Google \cite{arute2019nat}. 
Boson sampling \cite{aaronson2011computational} is an algorithm formulated specifically for photonics, which is realized by linear interference of highly ID single photons and sampling from the photon distribution. 
About 50 high-quality photons suffices reaching QA, although minor variations in the exact number may change as optimized classical algorithms are being developed. 
Figure \ref{fig:bosonsampling} quantifies the required quality of the photons for the boson sampling algorithm. 
So far a 20-photon boson sampling experiment with a QD source has been performed \cite{Wang2019prl}, while it has been shown that improved QD sources allow scaling further up and into the regime of QA \cite{uppu2020scienceadv}. 
An explicit photonics QA demonstration was very recently reported \cite{Zhong2020science} albeit in this case single-photon sources were not applied but rather squeezed light sources for realizing a Gaussian boson sampling algorithm.
Nonetheless, this experiment constitutes a very important milestone for photonic quantum computing in explicitly demonstrating a setup that can process and detect the many optical modes. 
Furthermore, the setup could be generalized to control also single photons from the optimized QD sources for an explicit QA demonstration of photonic qubit technology. 

What are the next steps beyond the QA demonstrations? 
This is another essential question. 
Indeed, the current QA simulators are not solving any relevant problems and hence to justify the huge experimental efforts required, it is essential that they constitute stepping stones towards addressing pertinent problems.
The present article highlights some of the opportunities identified for quantum photonics based on deterministic photon-emitter interfaces, including the route towards realizing them. 
The concept of QA, as discussed above, can be also formulated in a broader context than for quantum simulations. 
In this spirit, any protocol that exploits inherent quantum effects to realize applications that are not possible with classical resources could be referred to as QA. 
This entails applications such as device-independent quantum key distribution, quantum repeaters, and certain quantum sensing protocols, to mention a few examples. 
The break-down of these protocols in actual hardware architectures, including a thorough bench-marking of the hardware requirements for entering the QA regime, would constitute important guidelines for the quantum hardware development.


\section{Photonic building blocks}

The application of photon-emitter interfaces in quantum photonics technologies requires interfacing to additional functionalities. Quantum photonics is favorable since a modular scaling-up approach applies where fundamental building blocks of sufficiently high quality are combined into complex architectures. Furthermore, the hardware can be fabricated with advanced nanofabrication equipment that has been developed for classical photonics applications. Photonics technology offers high functional stability, mass productivity, and ultimate level of integration \cite{Sun2015nature}. Importantly, the quantum photonics applications are compatible with classical hardware development of PICs \cite{Bogaerts2020nat}, yet the performance requirements of quantum technology are pushing current boundaries,  in particular requiring ultra-low-loss operation. These improvements would lead to significant spill over of technology into the area of classical ``green IT technology'' \cite{Shaikh2015ieee} where the rapidly growing energy consumption of the internet is a concern \cite{Morley2018energy}. 
Quantum applications require low-loss performance due to the  ubiquitous ``no-cloning theorem'' \cite{Wootters1982nature} stating that quantum information cannot be amplified without noise penalties. As a consequence, the scalability of quantum photonics is intimately linked to the loss performance of all involved components. In the following we will briefly outline the functionality and performance of photonic devices required for scaling up.

\begin{figure}
    \centering
    \includegraphics[width=\columnwidth]{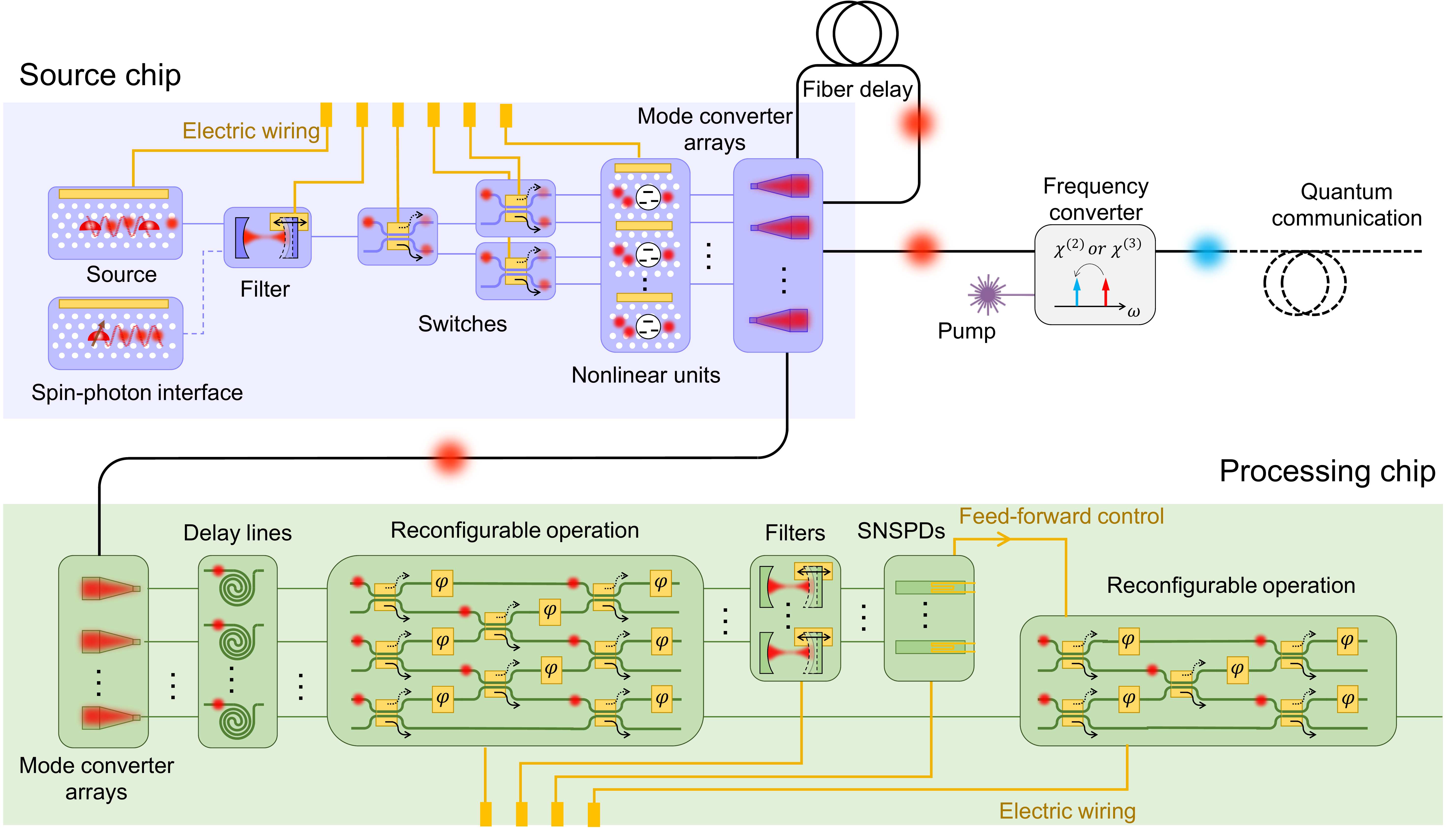}
    \caption{Illustration of basic functionalities required to construct a general-purpose quantum processor based in deterministic photon-emitter interfaces. A source chip comprises QD photon sources (single-photon sources and spin-based entanglement sources) together with spectral filters, photon routers, and nonlinear units. The prepared photonic resource is subsequently coupled off-chip with efficient mode converters for various applications either in quantum communication (fiber link) or quantum computing (processing chip). After the source chip optionally variable fiber delays and frequency conversion units are implemented. The source chip contains mode converters, on-chip optical delays, reconfigurable circuits to implement unitary optical transitions, filters, and detectors. Feed-forward from detection to the circuit is required as well. The various photonic modules are discussed in the main text. }
    \label{fig:photonic_modules}
\end{figure}

Figure \ref{fig:photonic_modules} outlines a vision for a general-purpose photonic quantum processor comprising photon sources, couplers, switches, converters, detectors, and more. A hybrid configuration  \cite{Elshaari2020np,kim2020optica} consisting of a source chip and a processing chip is sketched, which would be the most flexible approach with currently available technology based on material compatibility considerations. Long-term, the full integration of photonic quantum processors on a single chip can be envisioned.

The source chip is based on a direct bandgap semiconductor material, e.g. GaAs, hosting high quality QDs to produce photons. Photon-photon nonlinear interaction can be mediated by scattering off the QDs, which implements photonic two-qubit gates. Additional functionalities can be implemented on the source chip, e.g., filters to remove residual light from pumping or spin control pulses or switches to de-multiplex the photons. High efficiency mode converters are required to couple photons out of the source chip and into the processing chip. In between the two chips, frequency conversion modules could optionally be implemented, e.g., to convert the photon frequency to compensate variations between different QDs or to reach the telecom band as required for quantum communication. Furthermore, optical fiber delays can be inserted for proper timing of the photon stream. 

The processing chip would carry out the actual quantum operation, e.g., a quantum simulator algorithm, on the resource produced by the source chip. This generally requires a reconfigurable circuit in order to mutually interfere the photons combined with low-loss optical delay lines, filters and integrated single-photon detectors. In addition, fast feed-forward from the detectors to the reconfigurable circuit is essential  in many advanced applications \cite{zanin2020opex}. The detectors are preferably integrated in the processing chip, and superconducting nanowire single-photon detectors (SNSPDs) are very well suited for this.

Figure \ref{fig:photonic_modules} illustrates the building blocks for general-purpose photonics quantum-information processing.  For specific applications an actual chip design would need to be laid out. For the processing tasks, explicit photon-photon nonlinear interaction may be a major asset, therefore also more extended hybrid configurations can be envisioned involving more than the two chips illustrated in Fig.  \ref{fig:photonic_modules} or by active routing back and forth between the two chips. In the following, we will briefly outline the operational principles and specifications of the various building blocks of the proposed architectures. Our aim is not to exhaustively cover the vast amount of developments in integrated photonics, but rather to point to some opportunities for specific hardware that is compatible with the QD photon-emitter interfaces considered here.

\underline{Mode converters:} Routing photons efficiently in and out of the photonic chip is required  to realize hybrid architectures. The coupling efficiency from the chip and into a single-mode fiber is a relevant and quantitative figure-of-merit, although chip-to-chip efficiencies are important as well. Different approaches have been researched: end-fire coupling is favoured for its wide bandwidth and 86\% efficient coupling from an inverse tapered waveguide to a cleaved fiber has been realized \cite{cohen2013oe}. Another approach exploits surface grating for vertical outcoupling from the chip where special apodized grating couplers combined with a thin substrate metal mirror has reached 86\% coupling efficiency \cite{Ding2014ol}. Finally,  evanescent coupling between waveguides and tapered fibers have demonstrated impressive transfer efficiencies exceeding 95\% \cite{Tiecke2015optica}), although it remains a challenge to scale up this method to many fibers due to the demanding requirements in terms of alignment precision. For large-scale integrated quantum devices, a scalable approach is desirable, which favours the former two approaches.

\underline{Photonic switches}: Switches are key components in quantum photonics, since they enable  directing single photons into, e.g., different spatial modes, corresponding to single-qubit operations.
Essential switching figures-of-merits include the operation speed, switching contrast, insertion loss, and device footprint, and cryogenic compatibility is often a necessity.
Ultimately the overall switching speed (switch repetition and on/off time) is faster than the radiative emission time of the quantum emitter (i.e. sub-ns), since this would allow controlling each emitted photon. However, in many practical cases a much lower switching speed can be tolerated, since the photon source may not be operated at the highest possible internal repetition rate and/or switching of blocks containing, e.g., 10 photon pulses suffices. The switching of blocks instead of each individual photons only linearly decreases the count rate when de-multiplexing a deterministic single-photon source as opposed to optical loss that kicks in exponentially. Usable switching rates range from tens of MHz to several GHz, which are achievable, e.g., with  electro-optical devices\cite{Lenzini2017lpr} or nano-electro-mechanical devices\cite{Papon2019optica}.
These methods also allow controlling the splitting ratio, whereby arbitrary photonic qubits can be prepared.
As mentioned above, the switching loss budget is essential and various material platforms are considered featuring low-loss waveguides including silicon nitride (SiN) \cite{Bauters2011oe}, silicon (Si)\cite{Li2012oe}, and lithium niobate (LiNbO$_3$)\cite{Zhang2017optica}. The overall switch footprint, determined by the refractive index contrast and applied switching method, is essential for low-loss performance as well.
 Three electro-optical switches in LiNbO$_3$ have been integrated and operated at 80 MHz for realizing single-photon de-multiplexing of a QD into four modes \cite{Lenzini2017lpr}. Recent breakthroughs in thin-film LiNbO$_3$ technology have attracted much attention for building fast and low-loss switches\cite{wang2018nature} that are well suited for quantum applications. Furthermore, combining different materials provides a promising approach to boosting the electro-optic effects, leading to a lower footprint, as demonstrated on the Si platform\cite{he2019natphoton}.

Nano-opto-mechanical devices based on electrostatic or piezoelectric controls offer novel opportunities \cite{midolo2018nnano}. These devices feature small footprint and therefore ultra-low loss, since materials-induced electric-optical effects are not required. Furthermore, the capacitive nature of the actuation potentially leads to low electrical noise.  
  Switching speeds of up to 12 MHz have been achieved \cite{Haffner2019science} and single photons from a QD were routed with only 15 \% switch insertion loss \cite{Papon2019optica}. Furthermore, wafer-scale integration of 240 $\times$ 240 switching arrays have been realized \cite{Seok2019optica}, showing the exciting potential for scalability.

\underline{Optical filters}: Optical control pulses are generally required for operating photon-emitter interfaces, and on-chip optical filters allow rejecting, e.g., pump stray light or to  remove phonon sidebands for increased photon ID. Reconfigurability is essential in order to tune the filter to the individual QD frequency. Two approaches can be considered using either tunable high quality factor ($Q$) cavities \cite{elshaari2017natcom,elshaari2018nl,zhou2019chip}  or multi-layer gratings \cite{li2019oe}. 
Along with low insertion loss, ideal filters feature high extinction of stray light, tailored bandpass linewidths, wide-band tunability to cover the spectral inhomogeneity of QDs, and operation at cryogenic temperatures. Spectral filtering of QD single-photon sources has been realized using thermal and strain tuning with a hybrid approach \cite{elshaari2017natcom,elshaari2018nl} and based on nano-opto-mechanical tuning in monolithical GaAs devices \cite{zhou2019chip}. A quantitative benchmark is to realize $Q \approx 10^4$, which suffices for filtering of phonon sidebands.

\underline{Optical delays}: Processing of photonic quantum information typically requires optical delay lines, which can either be realized on-chip with low-loss optical waveguides/cavities or by routing photons off-chip and into an optical fiber delay. 
Photon propagation does generally not introduce decoherence (apart from residual loss) and therefore a low loss optical delay line controlled by an optical switch constitutes a practical quantum memory for photons. Ultra-low loss delay lines of up to 27 m (corresponding to 136 ns) have been realized on a SiN chip featuring $<$0.1 dB/m loss \cite{Bauters2011oe,lee2012natcom}. Such a delay, would suffice for coupling individual photons from deterministic chains of about hundred photons. For comparison, typical fiber loss is 3.5 dB/km at the current operation wavelength of QD sources (about 950 nm), which is improved to 0.18 dB/km at the telecom C-band.

\underline{Frequency conversion}: With current growth and fabrication methods, solid-state quantum emitters have limited tunability and inhomogeneity between emitters is an issue. Nonlinear frequency conversion may be implemented for overcoming these issues, and can conveniently translate the photon frequency all the way to the telecom C-band, as required for quantum communication applications. 
Frequency conversion applies a strong and tunable pump laser to bridge the energy difference between the initial and target photon frequencies using $\chi^{(2)}$ or $\chi^{(3)}$ nonlinear materials.  Frequency conversion of a QD source to the telecom C-band has been reported in an external  periodically-poled LiNbO$_3$ crystals leading to an end-to-end efficiency of $\approx$35\%  \cite{weber2019natnano}. The efficiency can likely be improved further, e.g., by engineering the coupling to the nonlinear crystal, since the internal conversion efficiency may reach near unity. It should be noted that the nonlinear conversion process could introduce decoherence thereby reducing ID of the converted photons, as has been reported recently and requires further attention and optimization \cite{morrison2021arx}.
Advances in thin-film LiNbO$_3$ \cite{wang2018optica} and modal phase-matching of GaAs waveguides \cite{chang2018lpr} hold promise of realizing on-chip $\chi^{(2)}$ nonlinear conversion.
Finally, the $\chi^{(3)}$ nonlinearity of integrated Si or SiN waveguides and cavities have been applied for frequency conversion of QD single-photon sources \cite{Singh2019optica}.

\underline{Single-photon detectors}: To scale up quantum photonics, all components need to be low-loss and therefore mutually compatible. This applies as well to the read-out of photonic quantum information. The recent decades have witnessed very important progress on single-photon detectors \cite{hadfield2009np}. Key specifications of single-photon detectors include: low timing-jitter, high speed operation, near-unity efficiency, low dark-count rates, and preferably compatibility with PIC technology. 
SNSPDs have emerged as a very promising technology matching all these requirements \cite{you2020nanophoton} reaching $\geq$98\% detection efficiency \cite{Reddy2020optica}, $>$1.5 GHz \cite{zhang2019ras} count rates, $<$10 dark counts/s  \cite{marsili2013np}, and $< 3$ ps  timing jitter \cite{Korzh2020natphoton}.
Furthermore, photon-number-resolving capabilities can be realized using arrayed SNSPDs with advanced configuration scheme \cite{zhu2018natphoton}. All this progress makes SNSPDs today a mature technology that can be readily implemented in the complex architectures considered here.

\underline{Reconfigurable photonic circuits}:  Advanced PICs can be fabricated in commercial foundries providing a very mature and flexible resource capable of processing large photonic resources. Quantum photonic PICs typically comprise an array of input waveguides containing photonic qubits that are subsequently coupled in a complex architecture of Mach-Zehnder interferometers. These circuits are reconfigurable by the use of thermo-optical transducers and can be scaled up to remarkable complexity. For example, a universal linear optics circuit was constructed based on 26 input waveguides and 88 Mach-Zehnder interferometers \cite{Harris2017natphoton}, although these systems typically use probabilistic photon sources  \cite{Carolan2015science}. Rooted in this technology, the perspective of the present manuscript is to add additional quantum photonics resources, notably deterministic single-photon and entanglement sources and quantum nonlinearity (cf. Fig. \ref{fig:photonic_modules}), to go beyond the paradigm of linear quantum optics for advanced applications. To this end, the maturity of PICs is a major asset of photonics as compared to other qubit technologies. With PICs technology, the ultimate scaling up to process thousands and millions of qubits can be foreseen, which is required for the long-term applications of fault-tolerant quantum computing \cite{Rudolph2017APLphoton}.


\section{Photonic quantum resources}
\begin{figure}
    \centering
    \includegraphics[width=\columnwidth]{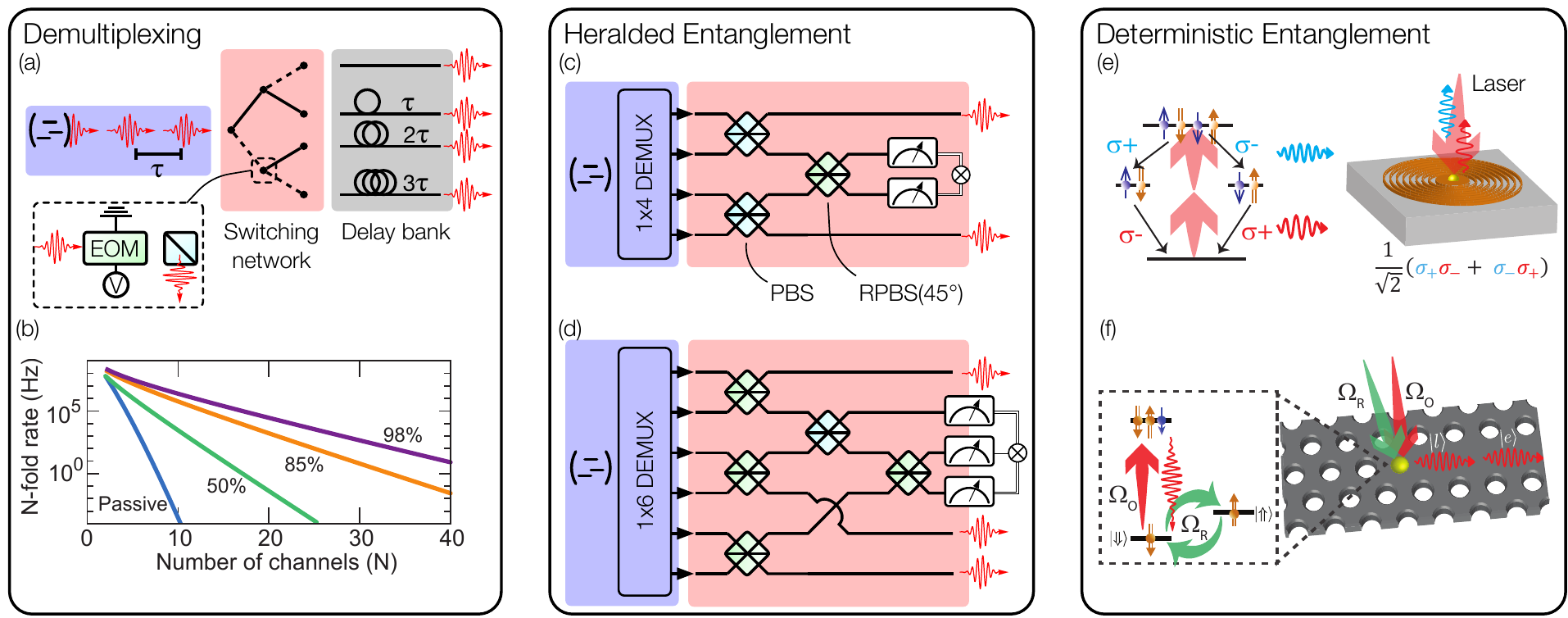}
    \caption{(a) Sketch of a demultiplexing setup that switches subsequently emitted photons from a deterministic source into separate spatial modes and compensate for the optical delays to produce N separate single-photon sources. Here $N=4$ is illustrated. (b) Rate of producing $N$ photons in an $N$-channel demultiplexing setup for a deterministic single-photon source (source efficiency $78\%$ operated at $1$ GHz repetition rate) for various values of loss per switching event and including also realistic fiber and mode-matching losses (total efficiency of $90\%$ \cite{Wang2019prl,Hummel2019apl}). (c), (d) Layout of photonic circuits for realizing heralded entangled Bell and three-photon GHZ states, respectively. (e) Level-structure of a biexciton cascaded decay producing a polarization entangled Bell state on demand that can be efficiently coupled out of rotation symmetric photonic nanostructure such as the indicated "Bull's eye grating". (f) Protocol for deterministic generation of a multi-photon cluster state by repeatedly exciting a QD that subsequently emits photons to the waveguide. By implementing coherent spin rotations entanglement is generated, where the qubit is encoded in either an early (e) or a late (l) time bin.} 
    \label{fig:resources}
\end{figure}

In this section, we will discuss the photonic quantum resources that can be realized with deterministic photon-emitter interfaces in conjunction with the building blocks considered in the previous section. 
By fully exploiting the quantum light-matter interface, a wide selection of high-fidelity quantum states can be prepared on demand, which testifies the flexibility of the approach being a major asset for scalability.

\underline{Multi-photon source:}
A deterministic single-photon source can be demultiplexed to realize multi-photon sources. 
A demultiplexing architecture is illustrated in Fig. \ref{fig:resources}(a): the deterministic train of photons is routed to different spatial modes by cascading electro-optical switches, and the photons mutual delays are compensated by inserting varying optical delays (e.g., fibers) in the  output modes. 
The scalability of this approach is ultimately determined by the residual switch and delay loss, cf. Fig. \ref{fig:resources}(b), while state-of-the-art QD sources are capable of delivering many high-quality qubits, as discussed previously. 
This highlights the general opportunity for QD sources: even few matter qubits can produce many high-fidelity photonic qubits that subsequently can be demultiplexed and processed. 
So far, demultiplexing of up to 20 simultaneous photons from a QD has been experimentally realized using bulk optical components \cite{Wang2019prl}, and the improved QD sources allow scaling into the QA regime, cf. Sec. II.  
Low-loss chip-integrated demultiplexing schemes could potentially scale the multi-photon sources even further, since the influence of out-coupling loss from the chip could be reduced.

\underline{Heralded entanglement sources}:
Based on highly ID multi-photon sources, more advanced entanglement sources can be synthesized by quantum interference. 
Heralding can be incorporated into the scheme, at the cost of additional photons, whereby entanglement can be generated on demand. 
Specific examples are two- and three-photon entanglement, exemplified by Bell and Greenberger-Horne-Zeilinger (GHZ) states. 
These are essential building blocks also for more advanced multi-photon entangled states; indeed it has been shown that three-photon GHZ states suffice for synthesizing a universal multi-photon cluster state by ballistic scattering in a linear optics circuit \cite{GimenoSegovia2015prl}. 
Figure \ref{fig:resources}(c) and (d) illustrate the linear optic circuits required to produce heralded Bell states and three-photon GHZ states starting from four and six photons, respectively \cite{Zhang2008pra, Varnava2008prl}. 
These protocols have been realized with probabilistic sources \cite{Barz2010natphot} thus with limited efficiency.
However, very recently heralded Bell-state generation was demonstrated with a deterministic QD source as well \cite{li2020arx}. 
Unfortunately, the linear-optics approach introduces an unavoidable effective loss, for instance, the Bell pair generation method of Fig. \ref{fig:resources}(b) succeeds with a heralding probability of $3/16$. 
Nonetheless entanglement generation rates exceeding MHz are within reach with deterministic QD sources, which would be an important step forward compared to the performance of probabilistic sources.

\underline{Deterministic Bell entanglement sources:}
An alternative route to entanglement generation exploits a QD radiative cascade.
Biexcitons consist of two electrons and two holes confined to a QD and can be deterministically prepared, thereby alleviating the need for heralding.
The radiative recombination of biexcitons have been proposed for Bell state generation \cite{Benson2000prl}, cf. Fig. \ref{fig:resources}(e). 
Here the presence of two indistinguishable decay paths imply that entanglement is generated provided that the intrinsic QD fine-structure splitting can be tuned to zero.
The efficiency of the entanglement source can be boosted with photonic nanostructures as well.
However in the present implementation, in-plane rotationally-symmetric structures are required to retain the polarization symmetry required for polarization entanglement.
Figure \ref{fig:resources}(e) shows a "Bull's eye grating" that has been successfully applied for Bell state generation \cite{liu2019natnano}. 
Biexciton cascaded emission can also be exploited for hyper-entanglement generation (entanglement in both time and polarization degrees of freedom) \cite{Prilmuller2018prl}. 
Such hyper-entangled states could enhance the channel capacity in quantum-communication protocols or enable deterministic entanglement purification \cite{Sheng2010pra}.  

\underline{Deterministic multi-photon entanglement sources:}
Introducing a spin in a QD leads to additional opportunities for the generation of entanglement. 
The coherent control of an electron's or hole's spin can be utilized to entangle subsequently emitted photons: the 'quantum knitting machine' \cite{gershoni2018celo}. 
A metastable spin ground state in a QD is obtained by tunneling in a single carrier (either an electron or a hole) in electrically gated devices; the corresponding level system is depicted in Fig. \ref{fig:resources}(f), where an external magnetic field is applied to align the spin and Zeeman-tune the energy levels. 
Spin-photon entanglement has been demonstrated \cite{gao2012nature}, which in combination with a repeated and alternating sequence of spin rotations and photon emissions, has led to the explicit demonstration of three-qubit entanglement \cite{Schwartz2016science}. 
It has been an open question how these encouraging results can be scaled up in future experiments given the physical imperfections of the photon-emitter interfaces. 
To this end, it was predicted that for realistic physical parameters, QDs in nanophotonic waveguides may generate long (i.e. $>10$) multi-photon cluster states with infidelity per photon of only $1.6 \%$ \cite{tiurev2020arx}, by implementing a particularly favorable time-bin encoding protocol, cf. layout of Fig.\ref{fig:resources}(f) . 
Excitingly the fidelity of such multi-photon entanglement states are reaching the demanding requirements for measurement based fault-tolerant quantum computing. 
It will be exciting to witness in the future whether such sources can break new ground for photonic quantum simulators and advanced quantum communication protocols possibly even without reaching the threshold of fault tolerance. 

\underline{Higher-dimensional photonic cluster states:}
For the most advanced quantum photonics applications, notably for measurement-based quantum computing \cite{Briegel2009natphys}, photon entanglement along a one-dimensional string is not sufficient. 
Rather two- or three-dimensional entangled clusters are required. 
Such higher-dimensional cluster states can be synthesized by linear optic fusion gates \cite{GimenoSegovia2015prl}, however at the cost of reduced efficiency and they require a vast amount of ancillary photons. 
Deterministic sources can be developed as well, either by coherently coupling two quantum emitters hosting spins \cite{Economou2010prl} or by routing back a one-dimensional photon string in real-time to the spin to create entanglement links beyond the nearest-neighbour \cite{pichler2017pnas}. 
Coupled QDs can be realized either via tunnel coupling or optical coupling, as discussed previously. 
However, an all-optical spin-spin gate may be realized as well by placing the QDs in each separate arm of an interferometer and sending a single photon through. 
Heralded by the observation of the photon in one output mode from the interferometer, a spin-spin gate operation can be realized \cite{Mahmoodian2016prl}. 
This gate can be implemented with near-unity fidelity and success probability, which holds in the limit where the $\beta$-factor of the photon-emitter coupling approaches unity.

\underline{Nonlinear quantum optics with photon-emitter interfaces:}
The deterministic photon-emitter interface can also be exploited as a giant photon nonlinearity leading to novel opportunities. 
A single quantum emitter can only scatter a single photon at a time, and if a narrow-band (relative to the emitter linewidth) photon interacts coherently with a high $\beta$-factor emitter, the scattering probability approaches unity. 
As a consequence, strong photon-photon correlations will be introduced during the scattering process if a pulse contains two or more photons \cite{lejeannic2021prl}. 
By coherently controlling a spin in the quantum emitter, this allows a photonic switch controlled by a single spin \cite{javadi2018nn} and if the spin is coherently controlled, a Schrodinger cat state can be produced \cite{chang2007natphys}. 
Such a nonlinear interaction constitutes a non-Gaussian photonic operation.
Interestingly, non-Gaussian operations constitute the missing key functionality in quantum-information processing based on continuous variables \cite{Braunstein2005rmp} as opposed to the discrete qubit technology otherwise considered here. 
Hybrid discrete-continuous variable photonic quantum architectures remain an interesting future research direction.


\begin{figure}
    \centering
    \includegraphics[width=\columnwidth]{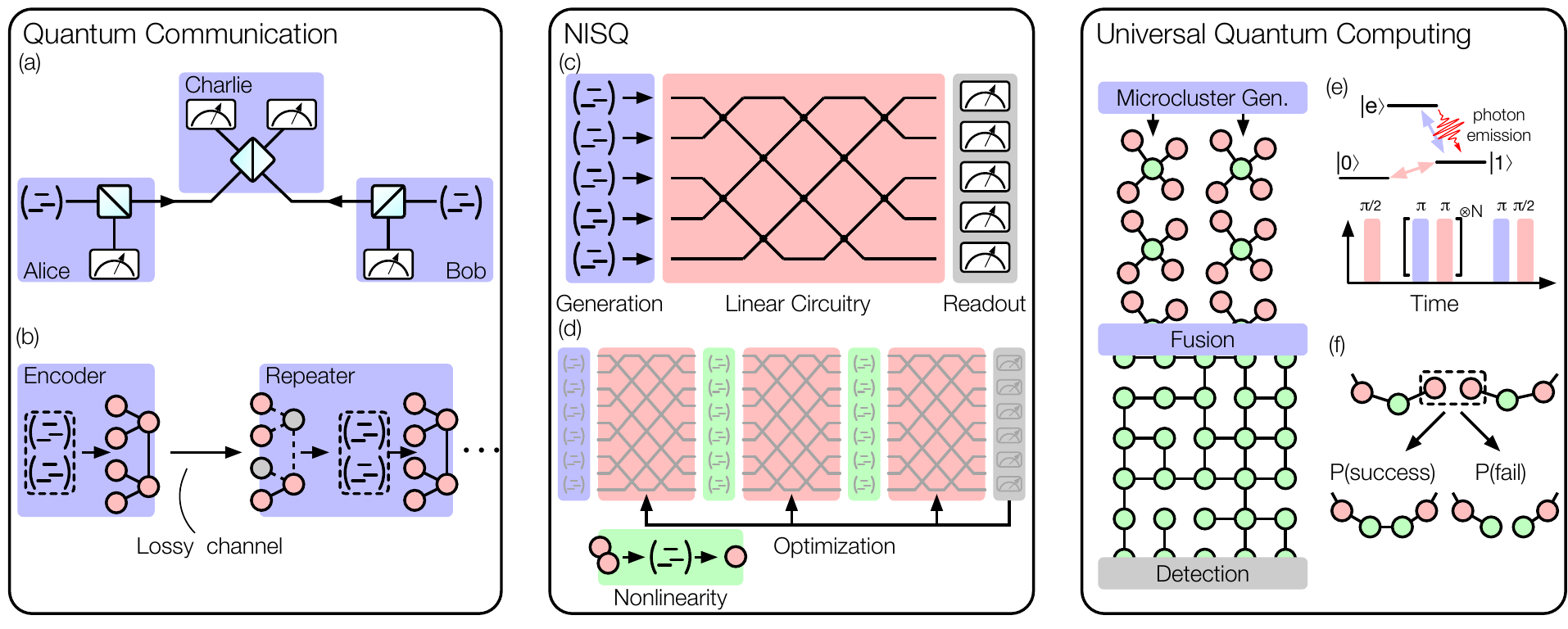}
    \caption{Applications of deterministic photon-emitter interfaces in quantum communication and quantum computing. (a) Generic quantum cryptography protocol for sending encrypted keys in single photons. (b) Operation principle of a one-way quantum repeater where a qubit is encoded non-locally in a photonic cluster state, sent through a lossy channel, and re-encoded at the next station. (c) Generic layout of a reconfigurable PIC that is fed by QD sources to realize a NISQ device. (d) Generic layout of a quantum photonics neural network is composed of sources, reconfigurable linear PICs, nonlinear interaction layers, and efficient detectors. (e) Illustration of small-scale photonic cluster state generation with QD sources. (f) Probabilistic fusion of small-scale cluster states into a percolated cluster state for quantum computing. }
    \label{fig:applications}
\end{figure}

\section{Applications}

Emerging quantum technology offers a plethora of novel applications in various areas.
Here we will focus on specific applications that photon-emitter interfaces seem particularly well suited for with no attempts of being exhaustive. 
Two main application areas will be discussed, respectively within quantum communication and photonic quantum computing. 

Quantum cryptography exploits encoded quantum information to distribute encrypted messages, and the security of the protocol is guaranteed by the laws of quantum mechanics \cite{Nicolas2002rmp}.
A quantum key can be distributed using a stream of single-photon qubits over long distances with the benefit of detecting any eavesdropping attempts on the transmission channel, cf. Fig. \ref{fig:applications}(a).
Quantum key distribution (QKD) could benefit from a deterministic single-photon source, as opposed to attenuated laser pulses that deliver single-photons probabilistically, since the ultimate bit rate is achieved when each communication pulse contains one and only one photon. 
High brightness QD single-photon sources are well suited for the task, but since QKD is a rather mature technology area, the higher costs associated with true single-photon sources as opposed to the cheaper alternatives may be an issue for most standard QKD protocols. 
Consequently, it is likely that deterministic sources will be of relevance in advanced QKD protocols offering ultimate security. 
Device-independent QKD is such a protocol and requires very efficient sources of highly indistinguishable single-photons for entanglement generation. 
The observation of the violation of the Bell inequality testifies that the system is protected not only against hacking attacks on the communication line but also against side-channel attacks on the receiver/sender hardware \cite{Vazirani2014prl}. 
High-quality deterministic single-photon sources have been proposed for a fully device-independent QKD implementation \cite{kolodynski2020quantum}, where the challenging requirements in terms of source efficiency and photon ID seem reachable with QD sources. 
Another related application of single-photon qubits is for the generation of a bit stream of random numbers. 
True random number generators have important applications in computing, e.g., in the context of Monte-Carlo simulation methods and also enables fundamental tests of quantum physics \cite{Herrero2017rmp}.
Physically randomness can be created  by reflecting single photons off a non-polarizing $50/50$ beam splitter to produce a binary bit stream of random numbers. 
Quantum random number generators can be made device-independent \cite{liu2018nature} in a similar manner as the QKD protocol discussed above.

Full-blown quantum computing is in principle possible based on single-photons and linear optics \cite{knill2001nature}. 
However, the resource requirements are staggering and additional hardware is required to make this approach more feasible. The present manuscript has introduced a few opportunities utilizing the nonlinear photon-emitter interface. It is interesting to consider whether specialized photonic quantum simulators can be developed for specific computing tasks within the current era of noisy intermediate scale quantum (NISQ) processors \cite{preskill2018quantum}. 
Measurement-based quantum computing protocols \cite{Briegel2009natphys} are generally well suited for photonics where the general idea is to produce a multi-photon entangled state up front and subsequently carry out single qubit measurements to implement the algorithm.
A promising direction is to tailor a multi-photon cluster state to a specific application or with a specific loss tolerance target \cite{Buterakos2017prx}, which could be significantly more resource efficient than starting with a universal cluster state potentially containing many redundant qubits. 

Quantum photonics is very well suited for simulating the dynamical evolution of complex quantum systems. 
Photons propagating through PICs emulate the physical system and the propagation depth of the PIC represents the evolved time, cf. Fig. \ref{fig:applications}(c). 
This is likely an area where photonics could offer quantum advantage in the near future utilizing current NISQ technology. 
So far, proof-of-concept quantum simulations of molecular vibrational dynamics have been carried out using probabilistic sources \cite{Sparrow2018nature}.
Such simulations could be scaled up further with deterministic single-photon and multi-photon sources, notably anharmonic vibrational effects would be of interest, which require a nonlinear interaction. 
Another emerging application area is the simulation of molecular dynamics problems, e.g., the dissociation of molecular bonds resulting from molecule-molecule interactions or the docking of a small molecule onto a larger host \cite{cao2019cr}. 
Despite being inherently quantum, such processes are today simulated by approximate molecular dynamics methods that rely on solving Newtonian equations of motion \cite{Henriksen2018book}. 
A photonic quantum simulator could be configured to treat such problems fully quantum mechanically and thereby testing the validity of existing methods. 
Precise simulations of vibrational dynamics and molecular docking are important in order to model complex protein folding problems, and a hybrid quantum/classical processor could be advantageous where a designated part of the problem is solved quantum mechanically while the rest can be approximated by classical means.
Importantly, computation of protein folding problems is a major challenge in drug discovery, and even modest computational advantages could be of major value and impact \cite{cao2018IBMjrd}.

Variational quantum algorithms (VQA) constitute another class of algorithms that are well suited for photonics due to the availability of flexible and reconfigurable PIC hardware. 
VQA requires only coherent quantum evolution of a very limited depth together with a classical algorithm that subsequently updates the quantum circuit before the next iteration. 
This makes NISQ hardware promising for VQA, and proof-of-concept photonics implementations determining molecular ground-state energies have  been reported \cite{Peruzzo2014natcomm}.
Quantum neural networks \cite{biamonte2017nature} provide another opportunity conveniently utilizing the reconfigurable PIC platform. 
The overall idea is to exploit the massive amount of information contained in large-scale quantum states as a novel resource for training algorithms. 
Quantum neural networks require access to nonlinearities and could be implemented in photonics via the deterministic photon-emitter interfaces, cf. Fig. \ref{fig:applications}(d) for an illustration of a  quantum photonics neural network \cite{steinbrecher2019npjqi}. 
Such a processor could be trained, e.g., to synthesize a desired multi-photon entangled state for a targeted measurement-based quantum algorithm.

The availability of multi-photon entanglement leads to additional opportunities also in quantum-communication applications. 
A general idea is to encode a qubit of information non-locally in a multi-photon entangled cluster, as opposed to using a single photon. 
Such multi-photon encoding makes the qubit more robust towards loss and errors.
The quantum communication `holy grail' is the quantum repeater \cite{Sangouard2011rmp}, which allows distributing quantum information over extended distances in the presence of unavoidable optical propagation loss. 
Ultimately the realization of a quantum repeater would pave the way for a quantum internet that could be used to scale-up quantum computers \cite{Kimble2008nature,Wehner2018science}. 
A long-lived quantum memory interfaced to the photonic links for efficient storage of photons would enable repeater architectures, however this is a challenging yet maturing research direction \cite{Heshami2016Jmodopt}. 
An alternative architecture is the `one-way quantum repeater' \cite{Fowler2010prl} that circumvents the requirement of a long-lived  quantum memory, and is therefore well suited for QD-based photonic hardware. 
The qubit is encoded non-locally in a cluster state at a transmitter station and entanglement distribution proceeds by directly transmitting the cluster state. 
In this case the redundancy of encoding in many photons implies that the encoded qubit is loss tolerant and can be re-encoded in a new photon cluster at the receiver station for further transmission, cf. Fig. \ref{fig:applications}(b). 
It has been proposed that coupled QDs can be configured to generate photonic cluster states suitable for quantum repeaters \cite{Buterakos2017prx} and based on that a blueprint of a QD-based one-way quantum repeater protocol was put forward and bench-marked \cite{borregaard2020prx}. 
This protocol was optimized and tailored to QD hardware such that only three QDs per repeater station were required and it was found to be realizable with experimentally feasible values of photon-emitter coupling, spin coherence, and spin-photon gate fidelity. 

Large-scale fault-tolerant quantum computing is the ultimate challenge for any quantum computing technology. 
It has been argued that photonic integration technology is a major \emph{raison d'\^{e}tre} for photonic quantum computing suggesting a real technological pathway to the daunting requirements for fault tolerance  \cite{Rudolph2017APLphoton}. Measurement-based quantum computing architectures \cite{raussendorf2001prl,walther2005nature} appear currently to be the most promising approach. 
It remains an open challenge whether the metrics of the photonic qubits can be of sufficient quality to reach fault tolerance. 
In the present manuscript, we have reviewed two approaches for creating the percolated large-scale photonic cluster state required for quantum computation: i) on-demand generation by coupled QDs in photonic nanostructures or ii) fusion of three-photon GHZ states. 
Approach i) has the advantage that the cluster state is delivered deterministically from the source, but is susceptible to imperfections of the QD sources leading to a decoherence of the state and hence limits the achievable cluster size. 
In approach ii), percolation of the cluster is done by linear optics, which does not introduce decoherence, but relies on probabilistic fusion of photons and therefore requires ancillary photons to boost the efficiency \cite{GimenoSegovia2015prl}.
Consequently an optimum strategy would likely be a combination of the two approaches where the QD sources are used for generating small-scale cluster states on-demand and linear optics subsequently allows growing the state bigger,  cf. Fig. \ref{fig:applications}(e) and (f).
Another opportunity would be to exploit the non-linear photon-photon interaction mediated by QDs to improve the photon fusion operation beyond the limitations set by linear optics. 
To this end, an explicit protocol for a Bell state analyzer has been put forward based on deterministic photon-emitter interfaces \cite{witthaut2012epl}.


\section{The road ahead}
Deterministic and coherent photon-emitter interfaces are now routinely realized in scalable solid-state devices, and we have summarized some of the near-term and long-term applications that this novel `photonic building block' could realize. 
The compatibility with a host of other photonic functionalities is essential, and we have highlighted the requirements and relevant specifications. 
Looking ahead, it is obvious that very serious engineering efforts are required in order to take the next step in this burgeoning technology area in order to tackle real-world problems. 
Indeed, in many cases the fundamental principles have been demonstrated for each device/functionality separately, but merging the building blocks together in advanced applications would introduce new tolerances in relation to fabrication yield and reproducibility, coupling loss, and cross-talk between devices. 
Excitingly, the high performance and thorough fundamental understanding of the building blocks now justify serious technology development, and likely we will see further quantum photonic hardware development gradually shifting towards industrial labs. 

While the all-solid-state platform based on QDs may have a number of appealing features, two main issues require additional attention: i) reducing emitter-emitter inhomogeneity and ii) coupling to a long-lived quantum memory. 
Although the protocols discussed in the present manuscript have been tailored to sidestep those `Achilles' heels' of QDs, it is clear that overcoming these obstacles would lead to an even more powerful and capable platform. 
The former challenge pertains primarily to the QD growth, and could be resolved if QDs were reproducibly synthesized with atomic precision at predetermined positions. 
The latter requires additional degrees of freedom for storage, and one promising candidate is to exploit coupling to the QD nuclear spins \cite{Gangloff2019science}, although this coupling may be difficult to control in present-day QDs due to asymmetric strain profiles. 
Alternatively, a hybrid approach may be pursued, where QDs are coupled to, e.g., ensembles of atoms or ions \cite{akopian2011np,Meyer2015prl} or ultra-long-lived opto-mechanical oscillators \cite{tsaturyan2017nn}. 
In these cases, efficient bandwidth and wavelength matching of the two systems is required, which could be pursued with nonlinear conversion, as was discussed previously. 

Hybrid interfacing to photonics may enable even more opportunities. 
In many qubit systems, e.g., based on spins or superconductors, qubit-qubit interactions beyond nearest neighbour are usually weak, which limits their scalability. 
An efficient quantum interface to photons would be a method for establishing such long-range interactions, and photon links have been proposed for scaling up ion-trap quantum computers following a modular approach \cite{Monroe2013science}. 
Such interfaces require proper quantum coherent transduction between the different qubit operation frequencies, e.g., transduction from microwaves to optical frequencies in the case of superconductor-photon coupling \cite{Mirhosseini2020nature}. A QD photon-emitter interface could be configured to implement such a transduction, e.g., by driving a tailored Raman transition in coherently coupled QDs \cite{Elfving2019pra}. 
Such a hybrid interface could lead to entirely new opportunities utilizing matter degrees of freedom for computation and photons as communication links. 
The availability of the coherent and deterministic photon-emitter interface today, the point of departure of the present manuscript, implies that such advanced hybrid interfaces are within reach. 
The ultimate dream of a large-scale quantum internet or a scaled-up quantum computer could be the outcome of such advancements.   

\begin{acknowledgments}
We thank Stefano Paesani for constructive comments on the manuscript. We gratefully acknowledge financial support from Danmarks Grundforskningsfond (DNRF 139, Hy-Q Center for
Hybrid Quantum Networks).
\end{acknowledgments}

\end{document}